%% file: main.tex
\documentclass[a4paper,submission]{eptcs}
\providecommand{\event}{PLACES 2010}
\usepackage{fancybox}
\usepackage{epsfig}
\usepackage{amsmath}
\usepackage{amssymb}
\usepackage{color,graphics,xcolor}
\usepackage{url}
\usepackage{graphicx}
\usepackage{color}
\usepackage{amsmath}
\usepackage{hyperref}
\usepackage{stmaryrd}
\usepackage{amsmath}
\usepackage{cmll}
\usepackage{mathpartir}
\usepackage{listings}
\title{A Modular Toolkit for Distributed Interactions}

\def\titlerunning{A Modular Toolkit for Distributed Interactions}

\author{
  Julien Lange \qquad Emilio Tuosto
  \email{ jlange@mcs.le.ac.uk \qquad emilio@mcs.le.ac.uk } 
  \institute{Department of Computer Science, University of Leicester, UK}
}

\def\authorrunning{J. Lange, E. Tuosto}

\input ifs
\input macro

\begin{document}

\maketitle

\begin{abstract}
  We discuss the design, architecture, and implementation of a toolkit
  which supports some theories for distributed interactions.
  The main design principles of our architecture are
  \emph{flexibility} and \emph{modularity}.
  %
  %
  Our main goal is to provide an easily extensible workbench to
  encompass current algorithms and incorporate future developments of
  the theories.
  With the help of some examples, we illustrate the main features of
  our toolkit.
\end{abstract}

\input{intro}


\input{background}

\section{Toolkit design and implementation}
\label{sec:framework}
\input{implementation}
\input{example}

\input{modularity}
\input{related}
\section{Conclusions}
\label{sec:plan}
\input{conc}




\label{sect:bib}
\bibliographystyle{eptcs}
\bibliography{stringdef,main}

\end{document}

%% file: ifs.tex
\newif\ifdraft
\draftfalse
\drafttrue

\newif\ifccalco
  \ccalcofalse
  \ccalcotrue
  
\newif\ifnat
  \natfalse
  \nattrue

\newif\iftr
\trfalse
\trtrue

\newif\ifllncs
  \llncsfalse
  \llncstrue

\newif\iffsttcs
  \fsttcstrue
  \fsttcsfalse

\newif\ifemi
  \emifalse
  \emitrue

\newif\ifthm
  \thmfalse
  \thmtrue

\newif\ifllncs
  \llncsfalse
  \llncstrue

\newif\ifarticle
  \articletrue
  \articlefalse

\newif\iftesi
  \tesitrue
  \tesifalse

\newif\ifmod
  \modtrue
  \modfalse

\newif\ifqapl
  \qapltrue
  \qaplfalse

\newif\ifqapl
  \qaplfalse
  \qapltrue

\newif\ifpar
  \parfalse
  \partrue

%% file: macro.tex
\newcommand{\Real}[1]{\mathrm{Real}}

\newcommand{\compile}[2]{\ifthenelse{\equal{#1}{yes}}{#2}{}}

\newcommand{\cf}[2]{
  \fontsize{#1}{#1}{\selectfont{#2}}
}
\ifemi
\usepackage{ulem}
\newcommand{\emi}[1]{{\marginpar{\cf{6}{{#1}}}}}
\newcommand{\emic}[2]{\par
  \definecolor{shadecolor}{rgb}{1,0.99,0.9}
  \fcolorbox{red}{shadecolor}{\parbox{\linewidth}{ 
      \color{gray}
      \begin{description}
      \item[{\color{blue} #2}]{\sf #1}
      \end{description}}}
}

\else
\newcommand{\emi}[1]{}
\newcommand{\emic}[2]{}

\fi

\ifemi

\else

\fi

\ifmod

\else

\fi

\ifqapl

\else

\fi

\newcommand{\proofend}{\mbox{$\Box$}}

\ifnat

\fi

\ifpar

\else

\fi

\newcommand{\strans}[1]{\xymatrix{\ar@{~>}[r]^{#1} &}}
\newcommand{\dstrans}[1]{\xymatrix{\ar@{~>}[d]^{#1} \\ \ }}

\iftesi

\else

\fi

\newcommand{\conf}[1]{\langle {#1} \rangle}

\iftesi

\else

\fi

\newcommand{\lab}[1]{\mbox{\small\tt #1}}

\ifarticle
\newtheorem{theorem}{Theorem}[section]
\newtheorem{definition}{Definition}[section] 
\newtheorem{proposition}{Proposition}[section] 
\newtheorem{lemma}{Lemma}[section]
\newtheorem{corollary}{Corollary}[lemma] 
\newtheorem{remark}{Remark}[section]
\newtheorem{observation}{Observation}[section]
\newtheorem{notation}{Notation}[section]
\newtheorem{example}{Example}[section]
\else
\ifthm
\else

\fi
\fi

\ifpar
\else

\fi

\iftesi

\else

\fi



%% file: intro.tex
\section{Introduction}
\label{sec:introduction}
With the emergence of distributed systems, communication has become
one of the most important elements of today's programming practise.
Nowadays, distributed applications typically build up from (existing)
components that are glued together (sometimes dynamically) to form
more complex pieces of software.
It is hence natural to model such applications as units of computation
interacting through suitable communication models.
An intricacy of communication-centred applications is that interactions
are distributed.
Here, the acceptation of distribution has to be taken in a very
general sense since interactions are physically and logically
distributed; as a matter of fact, components may run remotely and, for
instance, components may belong to different administrative domains.

In order to ensure predictable behaviours of communication-centred
applications, it is necessary that software development is based on
solid methodologies.
Besides the theoretical results that allow us to analyse systems and
prove their properties, it is also desirable to provide practitioners
with a set of tools to support them in addressing the most common
problems (e.g. avoiding synchronisation bugs).

In recent years, session types have appeared as an effective mathematical foundation for
designing and reasoning on distributed interactions.
For instance, dyadic session types \cite{Honda_languageprimitives}
have been proposed as a structuring method and a formal basis for the
verification of distributed interactions of two participants (e.g., in
client-server architectures).
Multiparty sessions~\cite{Honda_multipartyasynchronous} generalise session types
to support more than two participants; they have been used
in~\cite{dn10} to statically
compute upper bounds of the size of buffers used for asynchronous
communications in global interactions.
Moreover, dynamic multiparty sessions have been studied
in~\cite{Bruni_multiparty} and~\cite{Caires_conversationtypes}.
On top of multiparty sessions, in~\cite{bhty09}, a theory of
design-by-contract for distributed interactions has been introduced.
Basically, session types are extended with \emph{assertions} acting
as pre-/post-conditions or invariants of interactions.

%
Our main objective is to describe the design principles of the
architecture of a modular toolkit which puts in practise the theories
of distributed interactions based on session types. 
We have developed a toolkit that accommodates a few main requirements:
\begin{itemize}
\item Firstly, the toolkit provides a workbench for theoretical
studies so to permit ($i$) to experiment with potentially more
realistic examples and ($ii$) to possibly combine several of these
methodologies.
\item Secondly, our toolkit is easily extensible so to allow
researchers to explore new directions as the theory of distributed
interactions develops. For instance, the use of the tool revealed
some issues in the design of global interaction \`{a}-la~\cite{bhty09}
and triggered the design of additional algorithms to help the programmer design
correct distributed protocols~\cite{blt11}.
\item Finally, albeit being a prototype for research, our toolkit
shapes the basic implementations that can be used in more realistic
frameworks for the development of communication-centric software.
\end{itemize}

Arguably, most of the research around session types has been mainly
devoted to give a precise description of verification and
validation frameworks.
In fact, only very few and ad-hoc implementations have been developed
(e.g.,~\cite{Hu_sessionbased,Neubauer_animplementation,Pucella_haskellsession,Sackman_sessiontypes}).

The main contribution of this paper is the description of the design choice and the implementation of a modular toolkit that features the main algorithms necessary to analyse systems using theories based on session types. 
The most interesting features of our framework are illustrated by means of a few examples. The toolkit, its documentation, and a few examples are available at~\cite{toolsite}.


\noindent
\textbf{Synopsis} \S~\ref{sec:background} gives background information
and a motivating example.
\S~\ref{sec:framework} gives more details on the design principles of
our toolkit, its architecture and implementation.
\S~\ref{sec:example} gives illustrative examples of the tool's features.
\S~\ref{sec:modularity} highlights its main advantages.
\S~\ref{sec:related} compares our work to other implementation of ST.
\S~\ref{sec:plan} concludes and highlights our future plans.


%% file: background.tex
\section{Background and motivating example}\label{sec:background}
We briefly describe the distinguished aspects common to several
theories of distributed interactions.
The design principles of our toolkit hinge on some key elements of session
types that uniformly apply to several theoretical frameworks.
The key ingredients of the theories of distributed interactions based
on session types are described below.
\begin{description}
\item[Sessions] are sets of \emph{structured} interactions which
  correspond, for instance, to a complex communication protocol.
  Typically sessions are conceived as ``correct executions'' of
  a set of distributed interactions, namely those executions that
  run from the \emph{session initialisation} to its termination.
  The basic idea is that a computation consists of several concurrent
  sessions that involve some participants.
  A main concern is that participants acting in different sessions do
  not interfere.
  For instance, a desirable property to enforce is that a message sent
  in a session from $A$ and meant for $B$ is not received by another
  participant $C$; however, other relevant
  properties can be considered as, for instance, progress properties
  of sessions which guarantee that participants are not stuck because
  of communications errors.
\item[Interaction primitives] basically include communication
  mechanisms \`a-la $\pi$-calculus that deal with sessions as
  first-class values.
  Another kind of interaction primitives often present features
  a \textit{select/branch} mechanism which resembles a simplified form
  of method invocation.
  For instance, communication interaction and select/branch in the
  global calculus~\cite{Honda_multipartyasynchronous} notation are
  \[
  A \to B: k\conf{\texttt{sort}}
  \qquad \text{and} \qquad
  A \to B: k\{l_i:G_i\}_{i \in J}
  \]
  In the former, participant $A$ sends a message of type \texttt{sort}
  to $B$ on the channel $k$; in the latter $A$ selects one of the
  labels $l_i$ (sending it on $k$) and, correspondingly, $B$ executes
  its $i^{th}$ branch $G_i$.
  
  Communication primitives typically permit \emph{delegation}, namely
  the fact that sessions can be exchanged so to allow a process to
  delegate to another process the continuation of the computation.

\item[Typing disciplines] guarantee properties of computations.
  For example, in dyadic session types~\cite{Honda_languageprimitives} the
  \emph{duality} principle guarantees that, in a session, the actions
  of a participant have to be complemented by the other participant
  (or its delegates).
  Among the properties checked by type systems, \emph{progression}
  and some form of correctness properties are paramount.
  For instance, in~\cite{bhty09} a well-typed system is guaranteed
  to respect the contract specified by its assertions and, once projected,
  the program is guaranteed to be free from ``communication-errors''.

  Type systems are sometime subject to \emph{well-formedness}
  conditions.
  For instance, global types in~\cite{Honda_multipartyasynchronous}
  have to be linear in order to be projected onto \emph{local types}.
\end{description}

We illustrate some theoretical aspects with an example adapted
from~\cite{Honda_multipartyasynchronous} to the \emph{global
  assertions} in~\cite{bhty09}.
Intuitively global assertions may be thought of as global types
decorated with formulae of a (decidable) logic.
The following is a global assertion\footnote{
In this paper we deviate from the syntax adopted in~\cite{bhty09}
for assertions.
} specifying a protocol with two buyers ($B_1$ and $B_2$) and a seller ($S$).
The buyers $B_1$ and $B_2$ want to purchase a book from $S$ by
combining their money.
\newcommand{\assert}[1]{\llparenthesis\mathbf{assert}\ {#1}\rrparenthesis}
\newcommand{\eassert}{\llparenthesis \rrparenthesis}
\renewcommand{\lab}[1]{\textsf{#1}}
\[\begin{array}{llclclllr}
G = &  B_1 & \to &  S& : & s\conf{t: \texttt{string}} & \assert{t \neq ""}. & (1)
  \\[.5em]
  &S  & \to &  B_1 &: & b_1\conf{q: \texttt{int}} & \assert{q > 0}. & (2)
  \\[.5em]
 & B_1 & \to &  B_2 &: & b_2\conf{c:\texttt{int}} & \assert{0 < c \leq q}. & (3)
  \\[.5em]
 & B_2 & \to &  S &: & s\{\lab{ok}\ \eassert : D ,
  \lab{quit}\ \eassert : end\} & & (4)

\end{array}\]
The session $G$ above describes the interactions among $B_1$, $B_2$,
and $S$ after a session initialisation is performed\footnote{The
  session initialisation is not described in the global types or
  global assertions; it is an operation executed by the processes
  implementing the type $G$.}; such initialisation will assign a role
to each participant, namely each participant will act either as the
first buyer $B_1$, or the second one $B_2$, or else the seller $S$.
Each one of the interactions (1$\div$4) is decorated with an assertion
of the form $\assert{\phi}$ stating a condition $\phi$ on the
variables of the protocol ($\eassert$ abbreviates
$\assert{\lab{true}}$).
Basically, $G$ can be considered as a global type decorated with
logical formulae.

In (1), $B_1$ and $S$ interact (through $s$) and exchange the book
title $t$; the assertion decorating (1) states that $t$ is not the
empty string which means that $B_1$ guarantees $t \neq ""$ while $S$
relies on such assumption.
In (2), $S$ gives $B_1$ a quote $q$; similarly to (1), the assertion
$\assert{q > 0}$ constraints the price to a positive value and it
constitutes an obligation for $S$ and an assumption for $B_1$.
In (3), $B_1$ tells $B_2$ its non-negligible contribution $c$ to the
purchase (as $B_1$ guarantees $\assert{0 < c \leq q}$).
In the last step, $B_2$ may refuse (selecting label $\textsf{{\small
    quit}}$) or accept\footnote{For simplicity, it is not specified
  how $B_2$ takes the decision; this can easily be done with suitable
  assertions on $c$ and $q$.} the deal (selecting label
$\textsf{{\small ok}}$); in the former case the protocol just finishes,
otherwise it continues as:
\[
D = B_2 \to S: s\conf{a:\texttt{string}}\assert{a \neq ""}.
    S \to B_2: b_2\conf{d:\texttt{date}}\eassert
\]
namely $B_2$ and $S$
exchange delivery address and date.

Linearity is a (typically decidable) property ensuring that
communications on a common channel are ordered temporally.
Linear types can be \emph{projected} so to obtain the \emph{local
  types} for each participant.
In order to have effective algorithms, the theoretical framework
requires the decidability of the logic for expressing assertions as
well as the \emph{well-assertedness} of global assertions.
Informally, a global assertion is well-asserted when ($a$) each
possible choice a sender makes that satisfy the assertion of its
interaction is not making later senders unable to fulfil their
contracts (\emph{temporal satisfiability}) and ($b$) participants
state assertions only on known variables (\emph{history sensitivity}).

Well-asserted and linear global assertions can be projected, similarly to
global types, so to obtain local types, namely the interactions as
perceived from the point of view of each participant.
Unlike for global types though, projections of global assertions
must also ``split'' assertions in rely/guarantee propositions to be
assigned to each participant.
The projections of our example are:
\\[.5em]
\begin{tabular}{l|l}
  \begin{minipage}{.5\linewidth}
    \begin{tabbing}
      $pB_1$ = \= $s!\conf{t:\texttt{string}}\ \assert{t \neq ""} ;$\\
      \> $b_1?\conf{q:\texttt{int}} \ \assert{q > 0 \land t \neq "" };$\\
      \> $b_{2}!\conf{\texttt{c:int}}\ \assert{q > 0 \land 0 < c \leq q}$\\
    \end{tabbing}
  \end{minipage}
  &
  \begin{minipage}{.5\linewidth}
    \begin{tabbing}
      $pB_2$ = \= $b_2?\conf{\texttt{c:int}}\ \assert{\phi};$\\
      \> $s \oplus \{$\= $\textsf{{\small ok}}\ \eassert:$ \= $s!\conf{a:\texttt{string}}\ \assert{a \neq ""};$\\
      \> \> \> $b_2?\conf{d:\texttt{date}}\ \assert{ \phi \land a \neq ""},$ \\
      \> \> $\textsf{{\small quit}}\ \eassert: end\}$ \\
    \end{tabbing}
  \end{minipage}
  \\\hline
  \multicolumn{2}{c}{
    \begin{minipage}{\linewidth}
      \begin{tabbing}
        \\
        $pS$ =   \= $s?\conf{\texttt{string}}\ \assert{t \neq ""};$\\
        \> $b_1!\conf{q:\texttt{int}}\ \assert{q>0};$\\
        \> $s \& \{ $\= $\textsf{{\small ok}}\ \assert{ \psi }:
        s?\conf{a:\texttt{string}}\ \assert{ \psi  \land a \neq ""};
        b_2!\conf{d:\texttt{date}}\ \eassert ,$\\
        \> \>$\textsf{{\small quit}}\ \assert{ \psi }: end\}$
      \end{tabbing}
    \end{minipage}
  }
\end{tabular}
\\[.5em]
\begin{tabular}{lcl}
  where & $\phi = $&$\exists q:\lab{int},t:\lab{string} | \; 0 < c \leq q \; \land \; q > 0 \; \land \; t \neq ""$, and\\
        & $\psi = $&$\exists c:\lab{int} | \; 0 < c \leq q \; \land \; q > 0 \; \land \; t \neq ""$ \\
&&
\end{tabular}

The behavioural types $pB_1$, $pB_2$, and $pS$ above characterise
classes of processes that are ``well-behaved'' with respect to the
global interactions.
For instance, let us consider the process $cB_1$ below.
\begin{center}
{\tt \begin{tabular}{l l l}
    $cB_1$ = & $\bar{a}[2,3](s,b_1,b_2).$ & \textit{// Session initialisation} \\
    & $s!\conf{\text{``The art of computer programming''}};$ & \textit{// Send title to Seller}\\
    & $b_1?(\texttt{quote});$ & \textit{// Receive quote from Seller} \\
    & $b_{2}!\conf{\text{quote/2}}$ & \textit{// Send contribution to Buyer2}  \\
  \end{tabular}}  
\end{center}
The process $cB_1$ starts a session on $a$ declaring to act as the
first buyer of the global assertion $G$ above; this is done by the
action $\bar{a}[2,3](s,b_1,b_2)$ that will synchronise with the other
two participants (denoted by $2$ and $3$) using the session channels
$s$, $b_1$, and $b_2$.
It can be proved\footnote{We consider here a trivial process for
  simplicity; there are more complex cases where, for instance, the
  first buyer delegates the interactions with the second buyer to
  another process.} that $cB_1$ has type $pB_1$ which guarantees that
$cB_1$ has a correct interaction with any two other processes having
type $pB_2$ and $pS$.
The rest of the process is an instance of the type $pB_1$
detailing the behaviour of the first buyer.


%% file: implementation.tex
\subsection{Objectives}
The objective of this work is to describe the architecture and the
implementation of a modular toolkit implementing algorithms as those
described in \S~\ref{sec:background}.
The toolkit we developed supports the following
development methodology (see~\cite{savara} for a concrete realisation).
A team of software architects writes a global description of the
distributed interactions which specifies the intended behaviour of the
whole system.
The global description is checked and projected onto each participant.
Then, each part of the system is developed (possibly independently) by
a group of programmers.
Finally, the pieces of programs are checked, validated, and possibly monitored
during the execution.
This methodology is supported by the theories drafted in
\S~\ref{sec:background} whereby
\begin{enumerate}
\item global descriptions are given by global types and global
  assertions,
\item projections yield the parts of the systems to be developed, and
\item compliance of code with the specification is obtained by typing
  systems (to be matched against the projection).
\end{enumerate}
It is therefore possible to statically verify properties of
designs/implementations and to automatically generate monitors that
control the execution in untrusted environments.
Our main driver is that the architecture has to easily allow our
toolkit to be adapted to changes and advancements in the theories. For
instance, it has to consistently integrate the two (equivalent)
projection algorithms described in~\cite{bhty09}, or be parametric wrt
the logic used in the assertion predicates.
Note that our approach distinguishes itself from other works such
as~\cite{Hu_sessionbased,Neubauer_animplementation,Pucella_haskellsession,Sackman_sessiontypes}
by focusing on the tools accompanying the theories and not on the
integration of ST in a programming language.

\newcommand{\strm}[1]{\mbox{\sc stream #1}}
\subsection{Architecture}\label{sub:archi}
\begin{figure}[t]
\centering
\includegraphics[width=0.9\textwidth]{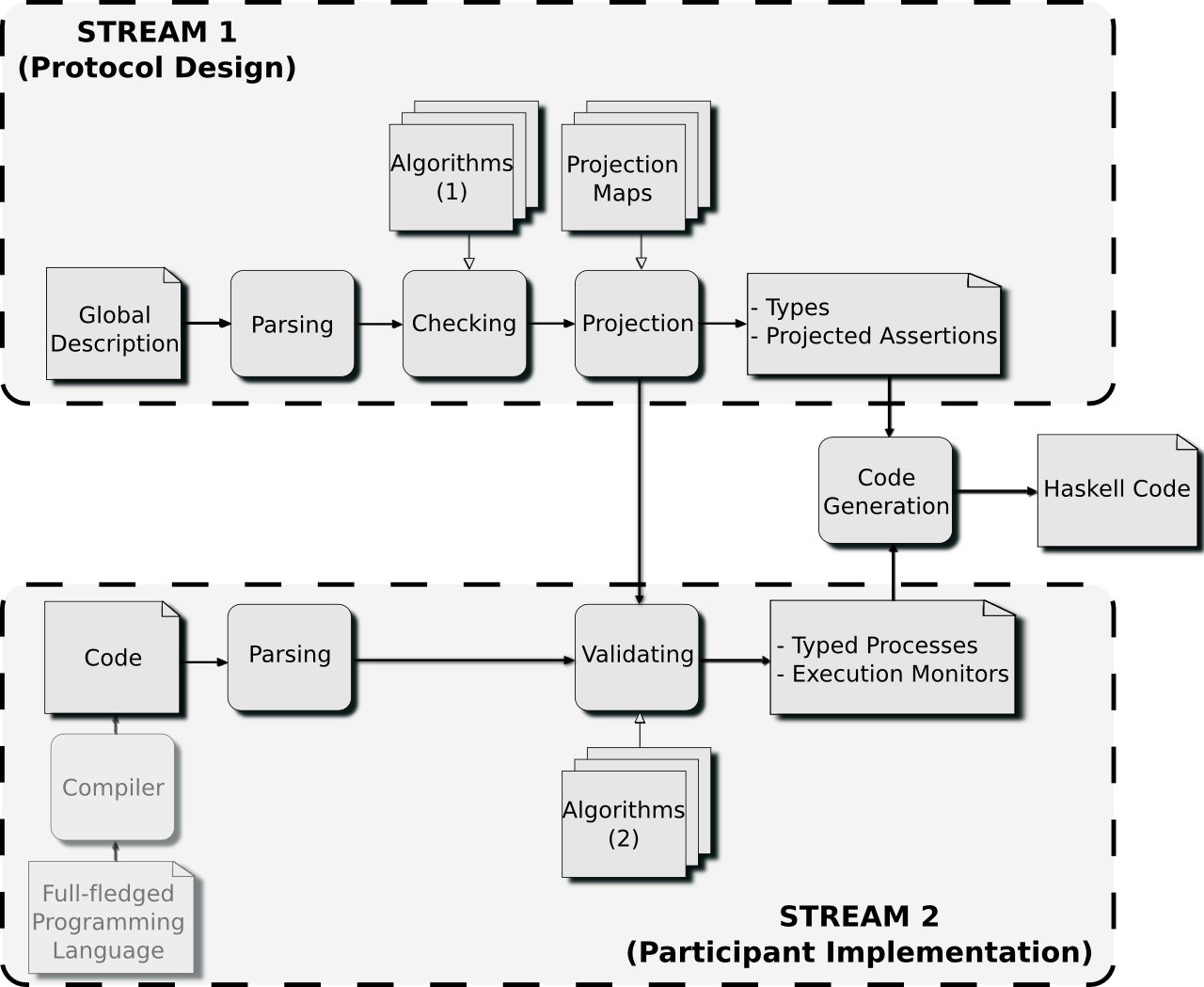}
\caption{Architecture}\label{fig:arch}
\end{figure}

The architecture of our workbench is illustrated in
Figure~\ref{fig:arch} and consists of two main streams, \strm 1 and
\strm 2; both streams' output are used for the \emph{code generation}
activity which combines behavioural types and processes to generate
safe Haskell code.
The two streams correspond to design of protocols, on the one hand,
and participants design and implementation, on the other hand.
The input of \strm 1 is a global description of the interactions while
the input of \strm 2 is the ``program code'' of each participant of the system, written in a dialect of the
$\pi$-calculus\footnote{The program code may be obtained by compiling
  programs written in full-fledged programming languages extended with
  session types like~\cite{Hu_sessionbased}.
  However, this feature is not yet available in the toolkit.
}.
In \strm 1, global descriptions are parsed, checked, and then
projected onto each participant.
In \strm 2, the code of participants is parsed, typed, and then
validated against the local types obtained in \strm 1 by
projecting the global interactions.
We give a walk-through of the architecture to illustrate the main
components of the toolkit.

\paragraph{\strm 1.} Taking a global description, such as $G$ in \S~\ref{sec:background}, a
parser constructs an abstract tree of the distributed interactions,
while interacting with the user if syntax errors are detected in the
description.
The checking module applies a series of algorithms (see $(1)$ in
Figure~\ref{fig:arch}) on the tree to check that some properties are
guaranteed.
At least, the following algorithms are executed.
\begin{itemize}
\item \emph{One-time unfolding}: recursive types are unfolded one time
  according to the equi-recursive view of recursive types.
  One-time unfolding is necessary before checking for linearity, whose
  definition relies on type equality.
\item \emph{Linearity check}: this is necessary to ensure that there
  is no races on the communication channels.
\item \emph{Well-assertedness}: if the global interactions are
  decorated with assertions, it has to be checked that they are
  well-asserted (cf. \S~\ref{sec:background}) in order to project them
  and obtain the local types as described in~\cite{bhty09}.
\end{itemize}
Each algorithm notifies the user in case the description does not
satisfy the properties, and accordingly, stop the execution of the
process.
If the global description is ``valid'', the projections (like $pB_1$,
$pB_2$, and $pS$ in \S~\ref{sec:background}) can be calculated.  This
is done according to the function defined in~\cite[\S 4]{bhty09} which
builds up on the projection operation in~\cite[\S
4.2]{Honda_multipartyasynchronous}.

\paragraph{\strm 2.}
A program code (written in a $\pi$-calculus-like language, such as $cB_1$ in \S~\ref{sec:background}) is parsed
to check for syntax errors and to build an abstract tree,
similarly to \strm 1.  Then, the following steps are applied on the tree.
\begin{itemize}
\item A typing algorithm infers the type of the processes according to a set of typing rules.
\item If the processes are asserted, a validator checks the interaction predicates for satisfiability.
\item The inferred types (and possibly the assertions) are compared with the projections.
\end{itemize}
As in \strm 1, each step of the stream notifies the user in case
errors are detected. In particular, mismatches between a participant's projection and its corresponding inferred types are explicitly output.


\paragraph{Code generation.}
In order to illustrate a practical use of our toolkit, we have
developed a prototypical translator which generates Haskell code from
the $\pi$-calculus-like code representing well-typed participants of a
global interaction.
On successful completion, \strm 1 and \strm 2 produce two outputs that
are compatible.
More precisely, safe Haskell code can be generated from the verified
$\pi$-calculus-like code given in \strm 2 and, possibly, execution
monitors\footnote{Monitors can be automatically generated from global
  assertions, but they are not yet part of the toolkit.}  can be
integrated from the projections computed in \strm 1.
This is possible due to the fact that inputs have passed all the
checks (i.e. processes are validated against the projections obtained
by the global interactions).

Further details on the code generation are given in
\S~\ref{sec:implementation}.

\subsection{Implementation}\label{sec:implementation}
An implementation of the toolkit has been developed in Haskell. Haskell has
been chosen because a functional language allows us to keep the implementation
close to the underlying theories and is more suitable for a large class
of algorithm in the toolkit.
For instance, the typing and projection algorithms can be
straightforwardly implemented by exploiting the pattern matching
featured by Haskell.
Moreover, Haskell provides a convenient means to build a modular
architecture; in fact, each component of Figure~\ref{fig:arch} is
implemented in a different module.
In addition, support of first-class functions allows for the
re-use of different functions in many different contexts (for instance,
to realise the parametricity of the toolkit wrt the assertion logic and
use the same typing algorithm for binary and multiparty sessions).

In the following, we discuss the main implementation details of the
current version of the toolkit.

\paragraph{Parsing.}%
Stable parsing tools are available for Haskell.
The parsers were built using Alex~\cite{alex} and Happy~\cite{happy}.
A basic attribute grammar takes care of checking conformance of the
code (e.g.\ basic type checking of the participants implementations is done at parsing time).
From the code input, it generates an abstract syntax tree (encoded in Haskell types) which is
then given to the next algorithm (\emph{linearity} check in the
multiparty case and \emph{typing} in the binary case).
The languages for global description and local processes accepted by
the toolkit are very similar to the ones defined
in~\cite{bhty09,Honda_multipartyasynchronous}. The main differences
are as follows.
\begin{itemize}
\item Session request $\overline{a}[2..n](\tilde{s})$ and session
  acceptance $a[2](\tilde{s})$ are respectively written as
  \[\begin{array}{l}
  \mbox{\texttt{init:a[$P_1 \ldots P_n$]($\tilde{s}$)}}
  \qquad \text{and} \qquad
  \mbox{\texttt{join:a[$P_i$]($\tilde{s}$)}}
\end{array}\]
where $P_i$ are participant identifiers. Note that the first participant identifier in the session request primitive is the initiator's.
\item The language adopted in the toolkit to represent processes
  requires that each branch construct is identified by a string which
  is then used as a prefix for the corresponding label selections;
  the syntax for the branch/selection constructs is respectively
  \[
  \mbox{\texttt{channel\&id\{\ldots\}}} \qquad \text{and} \qquad
  \mbox{\texttt{channel\$ [assertion] id.label}}
  \]
  This allows us to simplify the typing algorithm.
  Indeed, without such identifiers, it would be more complex to infer
  which branching construct a label is referring to.
  We illustrate this with an example; consider the following process
  \[\mbox{\tt
  \begin{tabular}{llllll}
    k & \& id   & \{ & t$_1:$ &  P$_1$; s \$ [-]  l$_1$ ; Q$_1$ & \\
       &        &    & t$_2:$ &  P$_2$; s \$ [-]  l$_2$ ; Q$_2$ & \\
       &        &    & t$_3:$ &  P$_3$; s \$ [-]  l$_3$ ; Q$_3$  &\}
     \end{tabular}}\]
   which branches on channel \texttt{k}; if the i$^{th}$ branch is
   selected, a label is sent on channel \texttt{s} after
   executing a process \texttt{P$_i$} (assuming \texttt{P$_i$} does not
   have interactions on \texttt{s}) and finally finishes with
   \texttt{Q$_i$}.

   Let the type of the processes interacting on  \texttt{s} be
   \[\mbox{\tt
     s  $\oplus $  \{   l$_1$ : T$_1$, \quad
     l$_2$ : T$_2$,  \quad
     l$_3$ : T$_2$ \} }\]
   To type channel \texttt{s} in the branches of the process, the
   algorithm needs a way to realise that all the labels \texttt{l$_i$}
   belong to the same branching construct. Since label selection can be done
   at any place in a process (e.g.\ in the branches of an \texttt{if-then-else}
   or a branching construct) and typing is done separately in the branches, one
   needs a way of gathering all the labels of a same group.
   Using an identifier for each branching construct and using it as prefix
   in label selections allows the algorithm to directly know which branching construct
   a \texttt{select} is referring to.
%
\item As in~\cite{bhty09}, recursive definitions for local processes take two kinds of parameters: expressions and communication channels. For example,
  \[
  {\tt
    mu \; t (e_1 \ldots e_n ; k_1 \ldots k_m) (p_1 : s_1 \ldots p_n : s_n; k'_1 \ldots k'_m).P
  }
  \]
is a recursive definition with $n$ ``value'' parameters and $m$ communication channel parameters. It is required that the type of each expression ${\tt e_i}$ matches ${\tt s_i}$.
The initialisation parameters ${\tt e_i}$ and ${\tt k_i}$ specify the initial value for the formal parameters  ${\tt p_i}$ and ${\tt k'_i}$. Each channel used in ${\tt P}$ must be one the ${\tt k'_i}$s.
\end{itemize}


\paragraph{Well-assertedness.}The logic used for the assertion is
based on the Presburger arithmetic~\cite{cooper72}, since a decidable logic was
necessary to develop an effective algorithm for checking the
well-assertedness condition on global assertions.
We have adopted a convenient API~\cite{presburgerhaskell} implementing the
Presburger arithmetic in Haskell.
The well-assertedness algorithm analyses which participants know which
variables to ensure \emph{history sensitivity} and tests the
satisfiability of the assertions as defined in~\cite{bhty09} to ensure
\emph{temporal satisfiability} (cf. \S~\ref{sec:background}).
A key part of the algorithm is the verification of temporal-satisfiability in
recursive definitions. It is implemented as follows.
Firstly, the algorithm checks whether the invariant is satisfied
by the assertions encountered previously, then it stores the formal parameters and the invariant of the recursion, and goes on with the verification of the continuation's temporal satisfiability.
When a recursive call is encountered, the satisfiability of the corresponding invariant
is tested in the new context (i.e.\ the actual parameters of the recursive call substitute
the formal parameters of the recursive definition). 

Note that because of the use of the API for the Presburger arithmetic, the
assertions one may write are quite limited, i.e.\ only predicates involving
integers and booleans are supported at the moment.
The well-assertedness algorithm was developed apart of the Presburger arithmetic API, to ease 
future changes in the underlying logic used in the assertions.
In case users do not want to assert their global description, all the assertions can be replaced
by \texttt{[-]} which stands for \texttt{True} in the language we defined.

\paragraph{Projection.} The projection of a global description is done participant-wise. The algorithm outputs a list of pairs (participant identifier and end-point type). If the description is asserted, then the projected assertions are computed at the same time. The output of this step is given to the typing algorithm.

\paragraph{Typing algorithm.}
The typing algorithm has been designed to be as flexible as
possible.
As an example, we use the same core algorithm for both binary and
multiparty session typing.
To make this possible, the typing algorithm is abstracted away from
two functions, the \emph{compatibility} and \emph{composition} operation
on type environments.
The former operation is used for testing compatibility between two
typings while the
composition operation is used to compose two typing
environments.\footnote{In the binary case, two typing
  environments $\Delta_0$ and $\Delta_1$ are \emph{compatible} if for every
  channel typed in $\Delta_0$ and $\Delta_1$, their types are the
  \emph{dual} of each other.
  In the multiparty case, $\Delta_0$ and $\Delta_1$ are \emph{compatible} if
  they type different participants for common channels.
  
  For dyadic session types, if a channel $k$ is
  typed in two compatible environments $\Delta_0$ and $\Delta_1$, then
  the type of $k$ in the \emph{composition} $\Delta_0 \circ \Delta_1$ becomes $\bot$ (i.e.\
  the interactions are internal).
  For multiparty session types, \emph{composition} consists, basically, of
  the union of typing environments.}
To type concurrent branches
(i.e.\ possibly representing different participants),
the algorithm first computes the type of each branch.
If the types of all branches can be successfully
obtained, the algorithm composes them using the parameterised
operations (provided that the obtained types are compatible).

The core of the algorithm consists of a depth-first traversal of the
abstract syntax tree. Each time a session initiation primitive
like \mbox{\texttt{init:a[$P_1 \ldots P_n$]($\tilde{k}$)}} or
\mbox{\texttt{join:a[$P_i$]($\tilde{k}$)}} is found, the algorithm
types the channels $\tilde{k}$ (called \emph{current channels})
in the rest of the tree according to the typing
rules specified in the theory.
Using pattern matching, it is straightforward to implement such rules
so to maintain a strong connection between theory and implementation.
%

An interesting rule is the one for typing recursive definitions.
When the algorithm encounters a recursive definition, e.g.\ 
  \[
  {\tt mu \; t (\tilde{e} ;\tilde{k}) (\tilde{s}:\tilde{S};\tilde{h}).P}
  \]
it first verifies that the types of the expressions in ${\tt \tilde{e}}$ match the
declaration of the formal parameters (i.e.\ ${\tt \tilde{S}}$).
The recursion variable, the formal parameters and the continuation are stored
in the environment.
Then, it continues the typing of {\tt P} where the current channels corresponding to the channels
in ${\tt \tilde{k}}$ are replaced by the corresponding channels in ${\tt \tilde{h}}$.

Once a recursive call is encountered, e.g.\
\[
  {\tt
    t (\tilde{e'} ;\tilde{k'})
  }
\]
the algorithm ensures, that  ${\tt P[\tilde{e'}/\tilde{s}][\tilde{k'}/\tilde{h}]}$
type checks, using the information previously stored in the environment.

In the multiparty case, when a session has been fully typed, its type
is compared to the corresponding projection. This is done using a
\emph{refinement} relation that allows the tool to accept processes
with types that specify a more refined behaviour that the one in the
projection.
For instance, a process may select less labels, weaken the predicate
for branching and reception, or strengthen the predicates for selection
and sending.

\paragraph{Code generation.}
The code generation is straightforward and exploits Haskell's
\texttt{Chan} objects for communication channels.
In Haskell, \texttt{Chan} is part of the concurrency libraries
provided and is an abstract type for unbounded FIFO channels.

Figure~\ref{fig:gencode} shows some example of generated code for
send, receive, branching and recursion respectively.
\begin{itemize}
\item \textsf{Send} is simply translated into a call to
  \texttt{writeChan} which writes a new value on the specified
  channel. All values are serialised\footnote{Note that serialisation
    in Haskell is supported through the inheritance of the
    \texttt{Show} and \texttt{Read} classes. Every type inheriting the
    \texttt{Show} class has to implement the function \texttt{show},
    from which a string representation of the object can be
    generated. Dually, a type which inherits \texttt{Read} defines a
    function \texttt{read} which can extract the data from the string
    representation. It is the case that \texttt{Show} and
    \texttt{Read} can be inherited for most of user defined
    types.} using \texttt{show} as channels accepts only one type of
  value per instance.
\item \textsf{Receive} is translated to a call to \texttt{readChan}, which
  reads data from a channel.
  When retrieving values from Haskell channels, it is necessary to
  cast back the string to its actual type (i.e.\ \texttt{read
    t'::(Int)}, in the example).
  This is needed as a Haskell compiler may not be able to infer the
  type of the value received.
  Remarkably, in our case the type is known in advance since the
  session was typed.
  The \texttt{let...in} construct of Haskell is quite useful since it
  allows us to bind the receive value to a new variable without having
  to take into account possible renaming.
  Indeed, nested \texttt{let...in} blocks declaring the same variable
  names are allowed and the scope of the binding corresponds to the
  one used in our language.
\item For branching blocks, one first reads one
label on the channel, then a \texttt{case} construct implements the
actual branching.
\item \textsf{Recursive} definition are translated into a \texttt{let...in} block where
  the type variable \texttt{t} is defined as a function with formal parameters
  \texttt{p$_1$, p$_2$, k$_1$} and \texttt{k$_2$} and body as the continuation of the recursive definition.
  In the \texttt{in} part the new function is instantiated with actual parameters \texttt{e$_1$}, \texttt{e$_2$}, \texttt{c$_1$} and \texttt{c$_2$}.
\end{itemize}
In order to make the execution of the generated code observable
it is possible to make the toolkit print information each time an action
such as send, receive, select and branch is invoked by a participant.

\begin{figure}[t]
\centering
{\tt {\small \begin{tabular}{l l|l}
{\bf Send:}    & s!(``The Art\ldots'')(t: string)[-]; & writeChan s ( show (``The Art\ldots'')); \\
&&\\
{\bf Receive:}        &  b1?(q: int)[-]; &  t' <- readChan b1;  \\
        &   (\ldots)  &  let (q) = read t'::(Int) \\
        &             &  in \\
        &             &  do \{(\ldots)\} \\
&&\\
{\bf Branch:}  & s\&id\{                                      & let brvarid' = read brvarid::String in   \\
                  &  [-] ok:  (\ldots)                                           & case brvarid' of \\ 
                  &  [-] quit: (\ldots)                        & ``idok'' -> do \{(\ldots)\} \\
                  & \}                      &  ``idquit'' -> do \{(\ldots)\} \\
&&\\
{\bf Recursion:} & mu t(e$_1$,e$_2$;c$_1$,c$_2$)(p$_1$:int,p$_2$:bool;k$_1$,k$_2$). & let t p$_1$ p$_2$ k$_1$ k$_2$ = \\
                 & (\ldots)                                                       & (\ldots) \\
                 &                                                                & in \\
                 &                                                                & t e$_1$ e$_2$ c$_1$ c$_2$
  \end{tabular}
}}
\caption{Example of generated code.}\label{fig:gencode}
\end{figure}

%% file: example.tex
\section{Examples}\label{sec:example}
In this section, we give two examples of distributed protocols. The first example
is taken from \S~\ref{sec:background} and illustrates how the toolkit can be used.
The second example gives a more realistic protocol, including a recursive definition,
which is verified by the toolkit.

\subsection{Buyer-Seller protocol} We start with the example given in Section~\ref{sec:background}.
Figure~\ref{fig:input} shows the input file given to the toolkit. The
first part (lines 1 - 9) represents the global description of the
two-buyer protocol ($G$ in \S~\ref{sec:background}), while the second
part (lines 11 - 34) gives an implementation of the participants
(Seller, Buyer1, and Buyer2) in our $\pi$-calculus dialect. These
processes are meant to be executed in parallel.

Buyer1 (lines 13 - 17, in Figure~\ref{fig:input}) sends the title of a book, receives its
price and, then, sends its contribution to Buyer2.  Buyer2 (lines 19 - 24)
receives the contribution that Buyer1 is willing to make. If
it is under 100, it confirms the sale to Seller and sends its
address.  Seller (lines 26 - 33) receives a book title, sends the
book's price to Buyer1 and then wait for Buyer2 to confirm, or not,
the sale.

The interactions are decorated with the assertions presented
in \S~\ref{sec:background}. However because of the limitation
imposed by the API for the Presburger arithmetic, it is currently not
possible to define assertion on strings, such as \mbox{\texttt{t} $\neq$ ""}.

In Figure~\ref{fig:input}, notice that Seller guarantees a stronger condition for the sending
on $b1$ (see line 28) compared to its counterpart in the global
description (line 4); this is made possible by the refinement
relation defined on local assertions (cf.~\cite{bhty09}).

\begin{figure}[t]
\centering
\begin{minipage}[c]{.9\linewidth}
\begin{lstlisting}
Global[buyerex]:
B1 -> S: s (t: string)[-].
S -> B1: b1 (q: int)[q > 0].
B1 -> B2: b2 (c: int)[0 < c and c <= q].
B2 -> S: s::id{
   [-] ok: B2 -> S: s (a: string)[-].
           S -> B2: b2 (d: date)[-].end,
   [-] quit: end
   }

Process:
{
init: buyerex[B1, B2, S](s, b1, b2).
      s!("The Art of Computer Programming")(t: string)[-];
      b1?(q: int)[q > 0];
      b2!(q)(c: int)[0 < c and c <= q];
      end   
|
join: buyerex[B2](s, b1, b2).
      b2?(c: int)[0 < c];
      if (c < 100)
      then s$ [-] id.ok; s!("my address")(a: string)[-];
                         b2?(d: date)[Exists q: int (0 < c and c <= q)];end
      else s$ [-] id.quit; end    
|
join: buyerex[S](s, b1, b2).
      s?(t: string)[-];
      b1!(99)(q: int)[q > 50];
      s&id{
       [-] ok:   s?(a: string)[-]; 
                 b2!(11/12/2010)(d: date)[-];end,
       [-] quit: end
      }
}
\end{lstlisting}
\end{minipage}
\caption{Buyer-Seller protocol.}\label{fig:input}
\end{figure}

When given the content of Figure~\ref{fig:input} as input, the implementation
outputs the text given in
Figure~\ref{fig:output} and generates Haskell code.
The toolkit first signals (lines 1- 2) that
the parsing was successful and the global description is well-asserted
(and linear). Then, the projections of the protocol (lines 6 - 18) on
each of the participants are given ($B1$, $B2$ and $S$ standing for Buyer1,
Buyer2 and Seller, respectively). Finally, the types of the processes,
prefixed by the session headings, are printed (lines 22 - 32), which
match the projections output before.  Notice that the predicates in
the projection of Seller (line 13) and in the type of its
implementation (line 29) are compatible since $q > 50 \implies q
> 0$.

When there is a mismatch between the projections and the types
inferred from the participants implementation, the toolkit shows the
problematic projection and inferred type.
For instance, if one changes the first interaction of Buyer1 (line 14 in Figure~\ref{fig:input}) to 
\texttt{s!(112)(t:int)[-];}, the tool outputs:
\begin{center}
\texttt{
  \begin{tabular}{ll}
    \multicolumn{2}{l}{Local type doesn't match projection for B1!}\\
Type:       & s!<t:int> [true];b1?<q:int>[q>0];b2!<c:int>[0<c \&\& c<=q];end \\
Projection: &s!<t:string>[true];b1?<q:int>[q>0];b2!<c:int>[0<c \&\& c<=q];end\\
\end{tabular}
}
\end{center}

In addition, if we set the book's price  to \texttt{0} in Seller, i.e.\
we change line 28  to
\texttt{b1!(0)(q:int)[q>50];} in Figure~\ref{fig:input}.
The tool signals that the assertion is not satisfiable:
\begin{center}
\texttt{[Typing-Send] Assertion not satisfiable: true => 0 > 50}.
\end{center}
Meaning that in the current assertion environment (which is empty, i.e.\ equals \texttt{true}), it is not
true that the sent value guarantees the assertion\footnote{In the near future, such error messages will
accompanied by a line number.}.

\begin{figure}[t]
\centering
\begin{lstlisting}
Parse Successful.
WellAsserted? [True]

Projections:
*buyerex:
[s!<t: string>[true];b1?<q: int>[q > 0];b2!<c: int>[0 < c && c <= q];end]@B1

[b2?<c: int>[Exist q: int st. (0 < c && c <= q && q > 0)];s${
  [true]ok: s!<a: string>[true];b2?<d: date>[Exist q: int st. (0 < c && c <= q && q > 0)];end,
  [true]quit: end
}]@B2

[s?<t: string>[true];b1!<q: int>[q > 0];s&{
  [Exist c: int st. (0 < c && c <= q && q > 0)]
      ok: s?<a: string>[Exist c: int st. (0 < c && c <= q && q > 0)];b2!<d: date>[true];end,
  [Exist c: int st. (0 < c && c <= q && q > 0)]
      quit: end
}]@S

Types: 
buyerex[s, b1, b2]: s!<t: string>[true];b1?<q: int>[q > 0];b2!<c: int>[0 < c && c <= q];end@B1
 
buyerex[s, b1, b2]: b2?<c: int>[0 < c];s${
  [true]ok: s!<a: string>[true];b2?<d: date>[Exist q: int st. (0 < c && c <= q)];end,
  [true]quit: end
}@B2
 
buyerex[s, b1, b2]: s?<t: string>[true];b1!<q: int>[q > 50];s&{
  [true]ok: s?<a: string>[true];b2!<d: date>[true];end,
  [true]quit: end
}@S
\end{lstlisting}
\caption{Output for Buyer-Seller.}\label{fig:output}
\end{figure}

\subsection{A guessing game protocol}
We use a protocol resembling  a simple game where a Generator (\texttt{G}) chooses a natural number which has to be discovered in less that 10 attempts by a Player (\texttt{P}), according to the hints given by a Server (\texttt{S}). The code representing the protocol and implementing the participants is given in Figure~\ref{fig:bigex}.
\begin{figure}[t]
\centering
\begin{lstlisting}
Global [a]:
G -> S: k (n: int)[n > 0].
P -> S: l (x: int)[x > 0].
mu rec (x,0) (r:int @ {P,S}, cpt: int @{P,S}) [r > 0 and cpt <= 10].
S -> P:h::id {
     [r > n and cpt < 10] less: P -> S: l (y: int)[y > 0]. rec(y,cpt+1),
     [r < n and cpt < 10] greater: P -> S: l (z: int)[z > 0]. rec(z,cpt+1),
     [r == n] win: end,
     [cpt >= 10] lose: end }

Process:
{
init: a[G,S,P](k,h,l).
      k!(15)(n: int)[n > 0]; end
|
join: a[S](k,h,l).
      k?(n: int)[n > 0];
      l?(x: int)[x > 0];
      mu rec (x,0;h,l) (r: int, cpt: int; hf,lf).
      if(cpt >= 10)
      then hf $ [cpt >= 10] id.lose; end
      else if (r > n)
           then hf $ [r > n and cpt < 10] id.less;
                lf?(y: int)[y > 0 and r > n and x > 0 and n > 0 and r > 0 and cpt < 10];
                rec(y,cpt+1;hf,lf)
           else if (r < n)
                then hf $ [r < n and cpt < 10] id.greater;
                     lf?(z: int)[z > 0 and r < n and x > 0 and n > 0 and r > 0 and cpt < 10];
                     rec(z,cpt+1;hf,lf)
                else hf $ [r == n] id.win; end
|
join: a[P](k,h,l).
      l!(11)(x: int)[x > 0];
      mu rec (x,0;h,l) (r: int, cpt:int; hf,lf).
      hf&id{
       [r > 1] less: lf!(r - 1)(y: int)[y > 0]; rec(y,cpt+1;hf,lf),
       [r >= 0] greater: lf!(r + 1)(z: int)[z > 0]; rec(z,cpt+1;hf,lf),
       [-] win: end,
       [-] lose: end }
}

\end{lstlisting}
\caption{Protocol description and implementation of a Guessing game.}\label{fig:bigex}
\end{figure}
The first part (lines 1-9) of Figure~\ref{fig:bigex} represents the global description of the game. First, \texttt{G} chooses a number $n>0$ and sends it to \texttt{S}. Then, \texttt{P} sends a first attempt (line 3) to \texttt{S}. The recursive definition has two parameters: $r$, the current attempt, initially assigned with value $x$, and \texttt{cpt} the attempt counter, initially assigned with value 0.
Depending on whether $r$ is less than, greater than or equal to $n$, \texttt{S} sends the corresponding label to \texttt{P}. If \texttt{P} guesses the correct number in less that 10 attempts, \texttt{P} wins and the session ends. Otherwise, the session ends after 10 attempts, and \texttt{P} looses.

The second part of Figure~\ref{fig:bigex} gives an implementation of each participants. In our example, \texttt{G} chooses always 15 for $n$ and \texttt{S} is faithful to the global description (i.e.\ it is not lying to \texttt{P}).
The participant \texttt{P} always starts with 11 as its first guess, then if \texttt{S} says \textit{less} (resp. \textit{greater}), \texttt{P} tries the number minus (resp. plus) one.
Remark that in both recursive definitions of participant \texttt{S} and \texttt{P}, two formal parameters are used for the communication channels, i.e. \texttt{hf} and \texttt{lf} (alternatively, one could have used the names \texttt{h} and \texttt{l} again).

\begin{figure}[t]
\centering
\begin{lstlisting}
Parse Successful.
WellAsserted? [True]
Projections:
*a:
[k!<n: int>[n>0];end]@G

[l!<x: int>[x>0];mu rec(x, 0){r: int, cpt: int}[r>0 && cpt<=10].h&{
[Exist n: int st. (r>n && cpt<10 && x>0 && n>0 && r>0 && cpt<=10)]
   less: l!<y: int>[y>0];rec(y, cpt+1),
[Exist n: int st. (r<n && cpt<10 && x>0 && n>0 && r>0 && cpt<=10)]
   greater: l!<z: int>[z>0];rec(z, cpt+1),
[Exist n: int st. (r = n && x>0 && n>0 && r>0 && cpt<=10)]win: end,
[Exist n: int st. (cpt >= 10 && x>0 && n>0 && r>0 && cpt<=10)]lose: end
}]@P

[k?<n: int>[n>0];l?<x: int>[x>0 && n>0];mu rec(x, 0){r: int, cpt: int}[r>0 && cpt<=10].h${
[r>n && cpt<10]
   less: l?<y: int>[y>0 && r>n && cpt<10 && x>0 && n>0 && r>0 && cpt<=10];rec(y, cpt+1),
[r<n && cpt<10]
   greater: l?<z: int>[z>0 && r<n && cpt<10 && x>0 && n>0 && r>0 && cpt<=10];rec(z, cpt+1),
[r = n]win: end,
[cpt >= 10]lose: end }]@S

Types: 
a[k, h, l]: k!<n: int>[n>0];end@G
 
a[k, h, l]: k?<n: int>[n>0];l?<x: int>[x>0];mu rec(x, 0){r: int, cpt: int}[true].h${
[cpt >= 10]lose: end,
[r>n && cpt<10]
   less: l?<y: int>[y>0 && r>n && x>0 && n>0 && r>0 && cpt<10];rec(y, cpt+1),
[r<n && cpt<10]
   greater: l?<z: int>[z>0 && r<n && x>0 && n>0 && r>0 && cpt<10];rec(z, cpt+1),
[r = n]win: end }@S
 
a[k, h, l]: l!<x: int>[x>0];mu rec(x, 0){r: int, cpt: int}[true].h&{
[r>1]  less: l!<y: int>[y>0];rec(y, cpt+1),
[r >= 0] greater: l!<z: int>[z>0];rec(z, cpt+1),
[true]   win: end,
[true]   lose: end }@P
\end{lstlisting}
\caption{Output for Guessing game.}\label{fig:bigexout}
\end{figure}

The code in Figure~\ref{fig:bigex} can be given as input to the toolkit, which then returns the content of Figure~\ref{fig:bigexout}.
Lines 4-22 of Figure~\ref{fig:bigexout} are the projections for participants \texttt{G}, \texttt{P}, and \texttt{S} computed from the global description of the guessing game. Remarkably, the projection for \texttt{G} is not a recursive type since \texttt{G} is not involved in the recursion.
The rest of Figure~\ref{fig:bigexout} consists of the local types for \texttt{G}, \texttt{S}, and \texttt{P}. These have been inferred from the participants implementation. Notice that the invariants of recursive definitions in local types are set to \texttt{true}. This is allowed since each local type must be a refinement of its corresponding projection, which is \emph{well-asserted} (i.e.\ it is guaranteed that the invariant is respected).
The toolkit also generates Haskell code implementing the participants, this can be compiled and ran as it is.


%% file: modularity.tex
\section{On featuring modularity}
\label{sec:modularity}
In this section we describe how modularity is featured in our
implementation. We mainly envisage four possible degrees of modularity
discussed below.

\noindent
\textbf{Notation.} \label{notation:itm} All inputs and outputs of the implementation (e.g. global assertions, projections, etc.) are encoded in Haskell data types that specify an abstract syntax of the supported languages. This allows to possibly support other notations than the ones originally considered. Notably, the implementation exhibits four data structures to/from which other languages can be translated:
\emph{global assertions},
\emph{end-point assertions} (projections),
\emph{$\pi$-calculus dialect} (participants implementation),
\emph{assertion logic}.

\noindent
\textbf{Languages.} \label{languages:itm} An important
  requirement of our modular approach, is that it has to feature the
  parametrisation of the implementation with respect to the languages
  used to describe the distributed interactions and the associated
  type systems. For instance, the theory described in \cite{bhty09}
  abstracts from the actual logical language used to express asserted
  interactions. Notably, depending on the chosen language, ad-hoc
  optimisations can be applied.

\noindent
\textbf{Algorithms.} \label{alg:itm} The tool consists of
  several algorithms that can be used in a modular way (i.e. the users
  will be able to choose which algorithms they need).  For instance,
  several algorithms are described in~\cite[\S 3.3]{bhty09} to check
  well-assertedness of assertions; in fact, depending on the adopted
  logic several formulae manipulation could be applied. Notably, the
  well-assertedness notion defined in~\cite{bhty09} could be replaced
  by equivalent ones which exploit optimisations on logical formulae.
  In this way, one could use the simple algorithms in theoretical
  experimentation on simple scenarios, while more efficient algorithms
  could be used when considering realistic cases.

\noindent
\textbf{Theory.} \label{theory:itm}  Since the toolkit is developed in a functional language, it allows the theory to be straightforwardly mapped into the programming language. This means that, most of the time, when one wants to change a rule or a definition this can be done by changing only a few lines of code.
We illustrate this with an example. The definition of the \emph{dependency relations} (\cite[\S 3.3]{Honda_multipartyasynchronous}) is translated as shown in Figure~\ref{fig:rules}.
In the conclusions of~\cite{Honda_multipartyasynchronous}, the authors comment the adaptation of the theory to support synchrony. Following their idea, this could be done by taking into account output-output dependencies between different names and adding a new dependency from output to input.
In our implementation this change could be implemented simply by a few modifications of the code in Figure~\ref{fig:rules}. In particular, we would relax the condition \texttt{k1 == k2} in \texttt{OO} and add a new \texttt{dep\_OI} function for output-input dependencies, similar to the other rules.

\begin{figure}[t]
  \centering
{\small
  \texttt{
    \begin{tabular}{r | l | l}
      \textbf{II} & $ n_1 <  n_2 $ and                          & dep\_ii :: Prefix -> Prefix -> Bool \\
                  &  $n_i = p_i \rightarrow p: k_i$ ($i$=1,2)   & dep\_ii (Prefix p1 p k1) (Prefix p2 q k2) \\
                  &                                            & | k1 /= k2 = (q == p) \\
                  &                                            & dep\_ii (Prefix p1 p k1) (Prefix p2 q k2) \\
                  &                                            & | k1 == k2 = (p1 == p2) \&\& (p == q) \\
                  &                                            & dep\_ii \_ \_ = False \\   
\hline
      \textbf{IO} & $n_1 < n_2$,                           & dep\_io :: Prefix -> Prefix -> Bool \\
                  & $n_1 = p_1 \rightarrow p: k_1$ and     & dep\_io (Prefix p1 p k1) (Prefix q p2 k2) \\
                  & $n_2 = p \rightarrow p_2:k_2 $         & | k1 /= k2 = (q == p) \\
                  &                                       & dep\_io \_ \_ = False \\
\hline
\textbf{OO} & $n_1 < n_2$,                                 & dep\_oo :: Prefix -> Prefix -> Bool \\
            &  $n_i = p \rightarrow p_i: k_i$ ($i$=1,2)    & dep\_oo (Prefix p p1 k1) (Prefix q p2 k2) \\
            &                                             & | k1 == k2 = (q == p)  \\
            &                                             & dep\_oo \_ \_ = False \\
\multicolumn{3}{r}{where each dep\_** $p_1$ $p_2$ assumes that $p_1 < p_2$.}\\
    \end{tabular}
  }
}
  \caption{Dependency relations implementation.}\label{fig:rules}
\end{figure}


%% file: related.tex
\section{Related work}\label{sec:related}
\paragraph{Implementations of session types.} A few other
implementations of the theories based on session types exist. We
describe them in the following and compare them with our work.

Hu et al.\ \cite{Hu_sessionbased} present an implementation of an extension of Java to support distributed programming based on binary session types. The implementation consists of three main components: an extension of the language to specify protocols, a pre-processor to translate the specification to Java and a runtime library which implements the communication channels and runtime checks.
Neubauer et al.\ \cite{Neubauer_animplementation} propose an encoding of session types in Haskell (in terms of type classes with functional dependencies). This implementation is quite limited, e.g.\ it is restricted to one channel.
Pucella et al.\ \cite{Pucella_haskellsession} proposes an implementation of session types in Haskell which infers protocols automatically. They also claim that their approach can be applied to other polymorphic, typed languages such as ML and Java.
Sackman et al.\ \cite{Sackman_sessiontypes} propose a full fledged
implementation of binary session types in Haskell in the form of an
API.
The type inference systems of~\cite{Pucella_haskellsession} and~\cite{Sackman_sessiontypes} are
based on Haskell's type system, i.e.\ they do not directly implement the typing rules defined
in the theories for session types.

Note that our work tackled the theories for \emph{multi}party sessions, while all the other implementations based on session types consider only two-party sessions.

Another difference between these works and ours is that we have
mainly focused on the tools accompanying the theories and not on the
integration of session types in a programming language.
The part of our work which can be directly compared with these
implementations is our Haskell code generator. However in its current
state, it is only a proof-of-concept of the usefulness of the
verification tool that are situated upstream in the toolkit. If we
were to generate code for more realistic applications, we would, e.g.,
use network connections instead of Haskell \texttt{Chan}. This would
notably require a good runtime library to support, for instance,
delegation of channels.


\paragraph{Other theories.}
We believe that our implementation could also encompass other theories
for distributed interaction. For instance, a sub-typing relation such
as the one defined by Gay et al.\ \cite{gh99} could be easily added to
the toolkit by using the \emph{refinement relation} in a new version
of the two parameterisable functions for the binary typing
algorithm. In particular, the \emph{compatibility} function should be
relaxed so to enable compatibility of types.
Similarly, to support \emph{union types} such as in~\cite{blc08}, it
would requires to add a few constructs to the accepted language (i.e.\
to allow the specification of types such as ``\texttt{int $\vee$
  real}'') and a new compatibility function which allows a type
containing a union of types to be compatible with one having one
element of it.

On the other hand, theories such as the one developed for the
\emph{conversation types}~\cite{Caires_conversationtypes} would be much
harder to integrate in our toolkit. For instance, the notion of
(nested) conversation would require to adapt the language and most of
the typing rules. In addition, the form of the types
in~\cite{Caires_conversationtypes} is quite different from most of the
other theories that we have focused on. However, provided a good
mapping between the theory of multiparty session
types~\cite{Honda_multipartyasynchronous} and conversation types in
terms of channels and conversations, we believe that an adapted
version of \emph{apartness} and \emph{merge} relations could be part
of the verifications done in the \emph{compatibility} function. These
two relations are used to test the compatibility of two conversation
types.  This would enable, to some extent, the verification of the
compatibility of participants without having a global description.

\paragraph{Applications.}
The toolkit aims to support a development methodology of
communication-centric software based on formal theories of distributed
interactions where global and local ``views'' are used to verify
properties of systems.
It is worth mentioning that a similar methodology has recently been
adopted in the SAVARA project~\cite{savara} where \emph{global} and
\emph{local} models are used in the development process to validate
requirements against implementations.
Noteworthy, SAVARA combines state of the art design techniques with
session types and provides an open environment where tools based on
formal theories can be integrated.
We are considering the integration of (part of) our toolkit in SAVARA.
In particular, our toolkit could be used to project the choreography
model onto individual services.
Namely, SAVARA uses WS-CDL~\cite{wscdl10} to represent the choreography
model, from which it can generates WS-BPEL~\cite{bpel06} implementations
of individual services and BPMN~\cite{bpmn07} diagrams that may be used
to guide the implementation.

We believe the integration of our tool in SAVARA is feasible and would
require the following three main components.
Firstly, a mapping from WS-CDL to global assertions. This should be
quite straightforward as the global types are very similar to
WS-CDL. However, the support for assertions may require more work as
it would demand an extension of WS-CDL with pre-/post-conditions on
messages.
Secondly, we need to translate the projections output by our tool to
BPEL, BPMN and/or another language, such as the ones used in SAVARA to
design/implement services.
Finally, we need some mechanisms to type check the conformance of
services against a choreography. This means that we need a tool which
translates the (partly) implemented services to a language compliant
with (the abstract syntax of) our $\pi$-calculus-like language.

Technically, these three mappings should be relatively easy to
implement as it amounts to transform an XML tree to Haskell data types
(and vice-versa). However, careful attention is needed when including
the assertions in the notations supported by SAVARA.


%% file: conc.tex
We have described the architecture and the main implementation aspects
of a toolkit for distributed interactions.
The distinguished design principles of the architecture are flexibility and
modularity, to meet the future changes in the theories underlying the toolkit.

The toolkit relies on Cooper's decision procedure
for the Presburger Arithmetic~\cite{cooper72}, which has a super-exponential complexity~\cite{oppen78}
on the size of the logic formulae. This means that the current version
of the algorithm might take a (very) long time to check protocol with very long
predicates.
However, all the algorithms of the toolkit itself have a polynomial complexity on the size
of the global descriptions and participant implementations. In addition,
the toolkit is designed in such a way that the underlying logic can be changed easily.
Therefore, if a more efficient logic API appears it can easily replace the current one.
To give an idea of the current efficiency of the toolkit, it takes more or less
one second to process the examples of Section~\ref{sec:example}, but a global
description decorated with a 80 line-long assertion (including quantifiers) takes about
11 minutes.\footnote{All the examples mentioned in this paper have been
ran in Ubuntu 10.04 with GHC 6.12.1, on an Intel Core 2 Duo @ 3.16 Ghz machine.}

Our toolkit is currently under development and we are considering
several ways of enhancing it. We are considering of using Haskell as
basis for the input languages.  For instance, one could use Haskell as
the language for expressions used, notably, in call such as
$s!("myString")$. In addition, Haskell types could be used as basis
for the types permitted by the tool.

Extending the input languages to Haskell might reveal a delicate
extension.  In fact, on the one hand this would provide the
possibility of expressing interesting predicates for the assertions,
on the other hand, global assertions require the logic
used to assert interactions to be decidable. In fact, increasing the
expressivity of the input language by allowing Haskell types might
compromise the decidability of the logic.

We intend to study the feasibility of a more realistic code
generator, which uses network connections (i.e.\ distributed Haskell)
instead of FIFO channels; and produces assertion monitor to ensure
runtime checking of assertions.


Another interesting implementation perspective would be to integrate
the algorithms featured by our toolkit in a full-fledged programming
language (e.g. Java, similarly to~\cite{Hu_sessionbased}). For
example, we conjecture that global assertions could be implemented in
two phases. Firstly, a language-independent part could take care of
the verification and validation tools which guarantee the good
behaviour of programs (i.e.\ the implementation of the toolkit
described here).  Secondly, a language-dependent part could extend a
programming language by developing an API which implements the
communication primitives (session initiation, value passing,
branch/select and delegation); while a translator to an abstract
language (such as the $\pi$-calculus-like language we use) links the
API to our toolkit (see faded boxes in Figure~\ref{fig:arch}).

\medskip

The toolkit turned out to be a remarkably useful tool to
identify which part of the theory relies on the programmers.
In particular, while designing tests for the toolkit, we noticed
that making a global description \emph{well-asserted} was often non-trivial.
This led us to design algorithms~\cite{blt11} which, if applicable, solve
well-assertedness problems automatically, and give indications to the 
programmers on where problems originate. We plan to add these algorithms
to the toolkit in the future.
